\documentclass[10pt]{article}

\usepackage{amsmath,amsfonts,amssymb}
\usepackage{paralist}
\usepackage{bm}

\newtheorem{theorem}{Theorem}
\newtheorem{proposition}[theorem]{Proposition}
\newtheorem{lemma}[theorem]{Lemma}

\newtheorem{definition}[theorem]{Definition}

\newtheorem{proof}{Proof}
\newcommand{\card}[1]{\left|#1\right|} 

\newcommand{\ccln}[1]{\langle#1\rangle}
\newcommand{\graph}[1]{#1^{\bullet}}
\newcommand{\np}{\mathrm{NP}} 
\newcommand{\p}{\mathrm{P}}
\newcommand{\csp}{\operatorname{CSP}}
\newcommand{\pol}{\operatorname{Pol}}
\newcommand{\core}{\operatorname{Cor}}
\newcommand{\cst}{\mathit{Cst}}
\newcommand{\sol}{\operatorname{Sol}}
\newcommand{\major}{\operatorname{major}}
\newcommand{\minor}{\operatorname{minor}}

\newcommand{\problemcsp}[3]%
{\begin{trivlist}
  \item[]%
    \textbf{Problem:} \textsc{#1}\\
    \emph{Input:} #2\\
    \emph{Question:} #3
  \end{trivlist}%
}

\begin{document}


\begin{center}
  \Large{Complexity of Homogeneous\\
    Co-Boolean Constraint Satisfaction Problems}
\end{center}

\begin{center}
  \textbf{Florian Richoux}\\ florian.richoux@polytechnique.edu\\
  Department of Mathematical Informatics, University of Tokyo
\end{center}

  
  \begin{abstract}
    Constraint Satisfaction Problems  ($\csp$) constitute a convenient
    way to capture many combinatorial problems.  The general $\csp$ is
    known  to  be $\np$-complete,  but  its  complexity  depends on  a
    template,  usually  a  set  of  relations,  upon  which  they  are
    constructed.  Following  this template, there  exist tractable and
    intractable instances of $\csp$s. It has been proved that for each
    $\csp$  problem over  a  given  set of  relations  there exists  a
    corresponding  $\csp$  problem  over  graphs  of  unary  functions
    belonging to the same complexity class. In this short note we show
    a dichotomy  theorem for every  finite domain~$D$ of  $\csp$ built
    upon  graphs  of  homogeneous  co-Boolean functions,  i.e.,  unary
    functions sharing the Boolean range $\{0, 1\} \subseteq D$.
  \end{abstract}

\section{Introduction}
\label{sec:intro}

Constraint Satisfaction Problems  ($\csp$) constitute a convenient and
uniform  framework  to  describe  many algorithmic  and  combinatorial
problems  from graph  theory,  artificial intelligence,  optimization,
computational molecular  biology, etc.  The general  $\csp$ problem is
well-known  to  be  $\np$-complete.   However,  we  can  consider  the
parametric version of the $\csp$ problem, denoted $\csp(S)$, where the
template~$S$ is a set of  allowed relations upon which any instance of
the problem  is constructed.  The goal  is to study  the complexity of
the  parametric  $\csp$, recognizing  the  conditions  allowing us  to
distinguish  between  tractable   and  intractable  instances  of  the
considered  problem, as well  as the  understanding of  the complexity
classes to  which these instances belong.  The  study of computational
complexity of constraint satisfaction problems was started by Schaefer
in   his  landmark   paper~\cite{Schaefer-78},  where   he  completely
characterized the complexity of Boolean $\csp$, distinguishing between
polynomial     and     $\np$-complete     instances.     Feder     and
Vardi~\cite{FederV-98} extended this  study to constraint satisfaction
problems over finite domains, for which they conjectured the existence
of a  Dichotomy Theorem.  So far,  this claim was proved  only for the
ternary  domain by  Bulatov~\cite{Bulatov-06}, exhibiting  an involved
Dichotomy  Theorem,   whereas  the  claim  remains   open  for  higher
cardinality domains.

A     fundamental     result     from     Feder,     Madelaine     and
Stewart~\cite{FederMS-04} shows  that for every  set of relations~$S$,
there  exists a  set~$F$ of  unary functions,  such that  the problems
$\csp(S)$  and   $\csp(\graph  F)$  are   polynomial-time  equivalent,
where~$\graph  F$ is  the set  of  the graphs  of functions  from~$F$.
Thus, $\csp$s over  unary functions are as powerful  as general $\csp$
problems.   Graphs  of  unary  functions  give us  a  very  structural
template which is really convenient to work with.

In this paper, we focus on templates built upon homogeneous co-Boolean
functions on a domain~$D$, that is, unary functions sharing a range of
size  two.  By  convention,  we  take the  range  $\{0, 1\}  \subseteq
D$. The goal of this paper  is more to present well-known results from
another  angle and  initiate  a new  way  to study  the complexity  of
$\csp(S)$  problems   rather  than  to   present  new  polynomial-time
algorithms for $\csp$.  The paper is organized as  follows.  The first
section  describes  general  notions  used  in this  paper.   Then  we
introduce  the   parametric  $\csp$   problem  in  general   and  more
specifically on graphs of homogeneous co-Boolean functions, as well as
some intermediary results.   In the last section, we  show a dichotomy
theorem for  every finite  domain~$D$ of $\csp$  built upon  graphs of
homogeneous  co-Boolean  functions.  The  paper  terminates with  some
concluding remarks.

\section{Preliminaries}
\label{sec:prelim}

Let $f\colon D  \to D$ be a  unary function over a finite  domain $D =
\{0, \ldots, n-1\}$.  This function~$f$ is called \emph{co-Boolean} if
the range  of~$f$, also named  the co-domain, is  of size 2.   In this
paper,   we   focus   on  \emph{homogeneous}   co-Boolean   functions,
\textit{i.e.}   co-Boolean functions sharing  the same  co-domain.  By
convention,  we  choose  $\{0,1\}  \subseteq  D$  to  be  this  shared
co-domain.   Since  in  this  short  note  we  deal  with  homogeneous
co-Boolean  functions  only,  we  can  simply  named  these  functions
"co-Boolean  functions"  without  any  confusions.   The  idea  behind
co-Boolean  functions  is  a  partition  of the  domain~$D$  into  two
disjoint sub-domains, where~$f$ acts as a characteristic function.

Since we study in this  paper only unary functions, each function will
be considered  to be unary  even if we  do not explicitly  mention its
arity.  We assume  that the domain~$D$ is ordered  by an arbitrary but
fixed  total order~$<$.   Without loss  of generality,  we  can assume
that~$<$ is the natural order $0 < 1 < \cdots < n-1$.  In other words,
the algebraic structure $(D; <)$ is a \emph{chain}.

The \emph{graph} of a function~$f$  is the binary relation $\graph f =
\{(x,f(x))  \mid x  \in  D\}$, extended  by  overloading to  a set  of
functions~$F$ as $\graph F = \{\graph f \mid f \in F\}$.

An $\ell$-ary relation~$R$ on a  domain~$D$ is a subset of $D^\ell$. A
relation~$R$  is  called  $0$-valid  ($1$-valid) if  it  contains  the
all-zeros tuple $(0 \cdots 0)$  (all-ones tuple $(1 \cdots 1$)). Given
a  tuple $t$ in  a $\ell$-ary  relation~$R$, we  denote by  $t[i]$ the
$i$-th coordinate  of $t$, with $1 \leq  i \leq \ell$.  We  say that a
relation~$R$ is \emph{closed} under (or \emph{preserved} by) a $k$-ary
operation~$p$, or that~$p$ is a \emph{polymorphism} of~$R$, if for any
choice of not necessarily distinct~$k$ tuples $t_1, \ldots, t_k \in R$
the following membership condition holds:
\begin{displaymath}
  \Bigl(
  p\bigl(t_{1}[1],\ldots,t_k[1]\bigr),\;
  p\bigl(t_{1}[2],\ldots,t_k[2]\bigr),\;
  \ldots,\;
  p\bigl(t_{1}[\ell],\ldots,t_k[\ell]\bigr)
  \Bigr)
  \in R,
\end{displaymath}
i.e.,  that  the new  tuple  constructed  coordinate-wise from  $t_1$,
\ldots, $t_k$ by  means of~$p$ belongs to~$R$.  We  denote by $\pol R$
the  set  of polymorphisms  of  a relation~$R$  and  by  $\pol S$  the
polymorphisms  of every relation~$R$  in~$S$.  Recall  that $\pol  S =
\bigcap_{R \in S} \pol R$.

In particular,  we need  to study the  closure under  four operations,
namely majority, minority, maximum,  and minimum.  Maximum and minimum
are   binary  operations,   satisfying   respectively  the   following
conditions for all elements $a,b \in D$:
\begin{displaymath}
  \begin{array}{rcl@{\qquad}rcl}
    \max(a,b)
    &=&
    \begin{cases}
      a & \text{if } a \geq b\\
      b & \text{otherwise}
    \end{cases},
    &
    \min(a,b)
    &=&
    \begin{cases}
      a & \text{if } a \leq b\\
      b & \text{otherwise}
    \end{cases}
  \end{array}
\end{displaymath}
Both  aforementioned  operations are  known  in  universal algebra  as
semi-lattice operations,  since they  correspond to the  operations of
join and meet. On the  Boolean domain $\{0,1\}$, the maximum operation
$\max(x,y)$  translates to  disjunction  $x \lor  y$  and the  minumum
operation  $\min(x,y)$ translates  to conjunction  $x \land  y$.  More
generally,  a \emph{semi-lattice operation}  is a  binary associative,
commutative and idempotent operation.  We say that a $k$-ary operation
$q\colon D^k \to  D$ is idempotent if the identity  $q(a, \ldots, a) =
a$ holds for every $a \in D$.

Majority and  minority are ternary  operations satisfying respectively
the following conditions for all elements $a,b \in D$:
\begin{displaymath}
  \begin{array}{rclclcl}
    \major(a,a,b) &=& \major(a,b,a) &=& \major(b,a,a) &=& a,\\
    \minor(a,a,b) &=& \minor(a,b,a) &=& \minor(b,a,a) &=& b.
  \end{array}
\end{displaymath}
It is clear that there  can be several majority and minority operation
on domains~$D$ of cardinality $\card D > 2$, whereas there is only one
majority  and  one minority  on  the  Boolean  domain $\{0,1\}$.   The
majority and  minority operations  on the Boolean  domain can  be also
expressed as  $\major(x,y,z) = (x  \lor y) \land  (y \lor z)  \land (z
\lor  x) =  (x  \land  y) \lor  (y  \land z)  \lor  (z  \land x)$  and
$\minor(x,y,z) = x+y+z \pmod{2}$.

There exists a \emph{pointwise partial order}~$\preceq$ on any $k$-ary
relation  $R \subseteq  D^k$ induced  by  the total  order~$<$ on  the
domain~$D$ defined as follows. Two  tuples $t$ and $t'$ from a $k$-ary
relation~$R$ satisfy $t  \preceq t'$ if $t[i] <  t'[i]$ holds for each
$i \in \{1, \ldots, k\}$. We write  $t \prec t'$ if $t \preceq t'$ and
$t \neq t'$.

A  \emph{constraint  language} is  a  set~$S$  of  relations over  the
domain~$D$.    Let~$X$   be   a   finite   set   of   variables.    An
\emph{$S$-constraint}  is  an application  $R(\vec  x)$  of a  $k$-ary
relation $R \in S$ to a  variable vector $\vec x = (x_1, \ldots, x_k)$
with $x_i \in X$ for  all~$i$.  An $S$-\emph{formula} is a conjunction
of $S$-constraints where variables can be existentially quantified. In
other words,  an $S$-formula  is a primitive  positive formula  of the
type $\exists  \vec y \; \bigwedge_{R  \in S} R(\vec x,  \vec y)$.  We
also  say  that  a  relation is  \emph{primitive  positive  definable}
from~$S$ if it is the set of models of an $S$-formula.  In the sequel,
we use the graph~$\graph F$ of functions~$F$ for the set~$S$.  In this
formalism, $\graph F$-constraints are written by means of equations of
the type $f(x) = y$ for a  function $f \in F$.  Note that we can write
an equation of the type $f(x) = g(y)$ for the expression $\exists z \;
f(x) = z \land  g(y) = z$ and $x = y$  for the expression $\exists z\;
f(z)  = x  \land f(z)  = y$.   A constraint  $R(x_1, \ldots,  x_k)$ is
satisfiable  if  there exists  an  interpretation  $I\colon  X \to  D$
satisfying the membership condition  $(I(x_1), \ldots, I(x_k)) \in R$.
A  conjunction  $R_1(\vec  x)  \land  \cdots  \land  R_k(\vec  x)$  is
satisfiable  if there  exists an  interpretation~$I$  satisfying every
constraint $R_i(\vec  x)$.  An $S$-formula $\varphi(\vec  x) = \exists
\vec y \; R_1(\vec x, \vec  y) \land \cdots \land R_k(\vec x, \vec y)$
is satisfiable  if the conjunction  $R_1(\vec x, \vec y)  \land \cdots
\land  R_k(\vec x,  \vec  y)$  is satisfiable.   We  write $I  \models
\varphi$  if the  interpretation~$I$ satisfies  the formula~$\varphi$.
The  set   of  \emph{models}   (or  \emph{solutions})  of   a  $k$-ary
formula~$\varphi$ is  the relation $\sol(\varphi(x_1,  \ldots, x_k)) =
\{(I(x_1), \ldots, I(x_k)) \mid I \models \varphi\}$.  If the identity
$\sol(\varphi)  = R$  holds  then we  say  that the  formula~$\varphi$
\emph{implements} the relation~$R$.

Given a relation  $R$ on a domain $D$ and  $p$ an endomorphism of~$R$,
we denote  by $p(R)$ the  relation $\{p(t[1]), \ldots, p(t[k])  \mid t
\in R\}$.  Similarly  for a set of relations~$S$,  we denote by $p(S)$
the set  of relations $\{p(R) \mid  R \in S\}$.  The  \emph{core} of a
constraint  language~$S$ is  the subset  $S_c \subseteq  S$  such that
every  endomorphism on~$S_c$  is an  automorphism.  Notice  that  if a
constraint language~$S$  is a  core then every  unary polymorphism~$f$
of~$S$  is bijective, i.e.,  $f$~is a  permutation on  the domain~$D$.
Observe that  all cores of  a constraint language~$S$  are isomorphic.
Thus,  we write  $\core S$  to  denote the  unique core  of~$S$ up  to
renaming.   Observe  also that  to  compute  a  core of  a  constraint
language~$S$, the polymorphism $p \in \pol S$ must be one of the unary
polymorphisms of~$S$ with the  smallest range applied on each relation
in~$S$.  Thus, the set $p(S)$ is a core of~$S$.

A \emph{relational clone}, also called  a \emph{co-clone}, is a set of
relations  closed  under  conjunction  (Cartesian  product),  variable
identification,  and   existential  quantification  (projection).  The
smallest co-clone containing a set of relations~$S$, denoted by~$\ccln
S$, is the set of relations primitive positive definable from~$S$.

\section{Constraint Satisfaction Problems}
\label{sec:csp}

In  general,  a  constraint  satisfaction problem  parametrized  by  a
constraint  language~$S$,  called  a  \emph{template}, is  defined  as
follows.

\problemcsp{$\csp(S)$}%
{An $S$-formula $\varphi(x_1, \ldots, x_k)$.}%
{Is~$\varphi$ satisfiable?}

\noindent%
In our context,  a \textbf{Co-Boolean Constraint Satisfaction Problem}
is a problem $\csp(\graph F)$ for a set of co-Boolean functions~$F$.

The  following theorem  allows us  to use  the algebraic  approach for
studying the  complexity of co-Boolean  CSPs.  We introduce it  in its
general form, for two arbitrary sets of relations.

\begin{theorem}[Jeavons~\cite{Jeavons-98}]\label{th:jeavons}
  Let~$S_1$  and~$S_2$  be  sets  of relations  over~$D$,  with  $S_1$
  finite.   If~$S_1  \subseteq   \ccln{S_2}$  holds  then  $\csp(S_1)$
  polynomial-time   many-one  reduces   to  $\csp(S_2)$,   denoted  by
  $\csp(S_1) \leq_p \csp(S_2)$.
\end{theorem}

To study the complexity of co-Boolean constraint satisfaction problems
$\csp(\graph  F)$  over  a  set  of co-Boolean  functions~$F$,  it  is
convenient   to   represent   the   set  of   graphs~$\graph   F$   in
\emph{$H$-normal form}.

\begin{definition}[\bm{$H$}-normal form]
  The~\bm{$H$}-\textbf{normal form} of the  of the set of functions $F
  = \{f_1,  \ldots, f_k\}$  is the $(k+1)$-ary  relation $F^H  = \{(d,
  f_1(d),  \ldots,  f_k(d))  \mid  d \in  D\}$.   The  \textbf{proper}
  $H$-normal  form is the  right-hand side  of~$F^H$, namely  $F_r^H =
  \{(f_1(d), \ldots, f_k(d)) \mid d \in D\}$.
\end{definition}

In other words, the $H$-normal form  of a set of functions $F = \{f_1,
\ldots, f_k\}$  is the  $k+1$-ary relation~$F^H$ which  is the  set of
solutions  of  a $\graph  F$-formula  $\varphi(x,  y_1, \ldots,  y_k)$
defined by the conjunction  $\bigwedge_{i \in \{1, \ldots, k\}} f_i(x)
= y_i$  and the  right-hand side  $F_r^H$ is the  set of  solutions of
$\exists x\;\varphi(x, y_1, \ldots, y_k)$.

We will  represent the relation~$F^H$ in  the form of  a matrix, whose
rows are  the tuples  of that relation.   The columns  of~$F^H_r$ then
represent the functions from~$F$, except for the first column which is
the enumeration of the domain $D$.  When we speak about~$F^H$, we call
the first column the left-hand  side and the other columns~$F^H_r$ the
right-hand side.

\begin{proposition}\label{prop:equiv}
  The  problems $\csp(\graph F)$  and $\csp(F^H)$  are polynomial-time
  equivalent.
\end{proposition}
\begin{proof}
  By definition we have the inclusion $F^H \subseteq \ccln{\graph F}$.
  Following  Theorem~\ref{th:jeavons} we  have that  $\csp(F^H) \leq_p
  \csp(\graph  F)$.   To recover  the  graph~$\graph f_i$  from~$F^H$,
  existentially  quantify all  coordinates of  $F^H(x_0,  x_1, \ldots,
  x_k)$  except  the  coordinates~$0$  and  $i$.   Hence  we  get  the
  inclusion  $\graph  F   \subseteq  \ccln{F^H}$.   This  implies  the
  reduction $\csp(\graph  F) \leq_p \csp(F^H)$,  concluding the proof.
\end{proof}

To classify  the complexity of $\csp(\graph  F)$ for any  set of unary
functions~$F$,  it  is  enough  to  classify  the  complexity  for~$F$
containing all unary constant functions.  This can be effectively done
by  means of  cores. Recall  that~$\core  \graph F$  denotes the  core
of~$\graph  F$.   Given  a  graph  of a  function~$\graph  f$  and  an
endomorphism~$\pi$ on $\graph f$, $\pi(\graph f)$ denotes the relation
$\{(\pi(d_i),\pi(d_j))  \mid (d_i,d_j)  \in  \graph f\}$.   Similarly,
given  the  graphs~$\graph F$,  $\pi(\graph  F)$  denotes  the set  of
relations $\{\pi(\graph  f) \mid  \graph f \in  \graph F\}$.   We need
first to prove that a core of graphs is a set of graphs.

\begin{lemma}\label{lemma:coregraph}
  Let~$F$ be a  set of functions.  There exists a  set of functions $G
  \subseteq F$ satisfying the equality $\core \graph F = \graph G$.
\end{lemma}
\begin{proof}
  Assume  that~$\graph  F$ is  not  a  core,  otherwise the  claim  is
  trivial.  Take  an unary polymorphism $p  \in \pol \graph  F$ with a
  smallest range  among all  unary polymorphisms of~$\graph  F$.  Then
  $p(\graph F)$ must  be a core of~$\graph F$,  implying the inclusion
  $\core \graph F \subseteq \graph F$, concluding the proof.  
  %
\end{proof}

We also need to show that the complexity of our $\csp$ problems do not
change if  we add all  constant functions to the  constraint language.
Let $\cst_D$ denote  the set of all unary  constant functions over the
domain~$D$.

\begin{lemma}\label{lem:corecst}
  For every set  of functions~$F$ on a domain~$D'$  there exists a set
  of  functions~$G$   on  a  domain   $D  \subseteq  D'$,   such  that
  $\csp(\graph  F)$  and  $\csp(\graph  G  \cup  \graph  \cst_D)$  are
  polynomial-time   equivalent.   More   specifically,   the  set   of
  functions~$G$  satisfies the equality  $\graph G  = \core  \graph F$
  and~$D$ is the domain of the constraint language $\core \graph F$.
\end{lemma}
\begin{proof}
  The  proof   is  a   direct  consequence  of   Theorems~4.4  and~4.7
  in~\cite{BulatovJK-05}.    Theorem~4.4  shows   that   the  problems
  $\csp(\graph  F)$  and $\csp(\core  \graph  F)$ are  polynomial-time
  equivalent. Theorem~4.7  shows that the  problems $\csp(\core \graph
  F)$   and   $\csp(\core   \graph   F  \cup   \graph   \cst_D)$   are
  polynomial-time equivalent.
  %
\end{proof}

According  to  the aforementioned  Lemmas,  we  assume  in the  sequel
that~$\graph F$  is always  a core for  any set of  functions~$F$, and
that~$F$ contains all unary constant functions, in particular $\bot(x)
=  0$  and   $\top(x)  =  1$  for  all  $x   \in  D$.   Therefore  the
relation~$F^H_r$ cannot be closed  under any constant operation, since
there exists no constant row in the matrix~$F^H_r$.

\section{Dichotomy Theorem}
\label{sec:dicho}

We present  a complete  characterization of complexity  for co-Boolean
constraint  satisfaction problems.   Recall that~$F^H_r$  is  a $\card
F$-ary Boolean relation, also considered as a $\card D \times \card F$
Boolean matrix.

First, let  us analyze the  case when the Boolean  relation~$F^H_r$ is
not closed under  any of the four particular  Boolean operations. This
case behaves similarly to the case observed in~\cite{Schaefer-78}.

\begin{proposition}\label{prop:nonclosed}
  Let~$F$ be a set of co-Boolean functions. If the relation~$F^H_r$ is
  closed neither  under majority,  nor minority, nor  conjunction, nor
  disjunction, then $\csp(\graph F)$ is $\np$-complete.
\end{proposition}
\begin{proof}
  Recall  that   the  relation~$F^H_r$  can  be   implemented  by  the
  constraint $\exists x_0 \;  F^H(x_0, x_1, \ldots, x_k)$ where $\card
  F  = k$,  and  remember  we assume  that  $F$~contains the  constant
  functions~$\bot(x)=0$ and~$\top(x)=1$  for all $x \in  D$. Since the
  co-domain   of   each    function   from~$F$   is   $\{0,1\}$,   the
  relation~$F^H_r$ is Boolean.  The Boolean relation~$F^H_r$ cannot be
  $0$-valid   or   $1$-valid,    since~$F$   contains   the   constant
  functions~$\bot$    and~$\top$.     Moreover,    if   the    Boolean
  relation~$F^H_r$   is   not   closed   under   majority,   minority,
  conjunction,   or   disjunction,   then  according   to   Schaefer's
  result~\cite{Schaefer-78},    $\csp(F^H_r)$    is    $\np$-complete.
  Therefore by Proposition~\ref{prop:equiv},  $\csp(\graph F)$ must be
  $\np$-complete, too.
  %
\end{proof}

The cases left  to be analyzed are those  when~$F^H_r$ is closed under
majority, minority,  conjunction, or  disjunction.  The first  case is
still compatible with Schaefer's results in~\cite{Schaefer-78}.

\begin{proposition}\label{prop:majorminor}
  Let~$F$  be   a  set  of  co-Boolean  functions.    If  the  Boolean
  relation~$F^H_r$   is  closed  under   majority  or   minority  then
  $\csp(\graph F)$ is in~$\p$.
\end{proposition}
\begin{proof}
  We  only give the  proof for  majority, since  the minority  case is
  completely  analogous.  The left-hand  side of~$F^H$  represents the
  enumeration  of the domain~$D$  and therefore  it can  be seen  as a
  numbering   for  the   tuples  in   the   Boolean  relation~$F^H_r$.
  If~$F^H_r$  is closed  under the  majority operation,  then $\major$
  applied to  any three (not  necessarily distinct) tuples  $a,b,c \in
  F^H_r$ results in a tuple  $d \in F^H_r$, where $\major(a,b,c) = d$.
  If we relate the  tuples $a,b,c,d$ by their corresponding numberings
  $l(a)$, $l(b)$,  $l(c)$, $l(d)$ in  the left-hand side  of~$F^H$, we
  get an extension  of the majority operation on  the whole domain~$D$
  in the following  way.  It is clear that  the values $l(a)$, $l(b)$,
  $l(c)$ must  be distinct for  different rows of the  matrix~$F^H$. A
  majority operation $\major(x,y,z)$ can  assume an arbitrary value if
  the three  values substituted  for the variables  $x$, $y$,  $z$ are
  different.   Hence we define  $\major(l(a), l(b),  l(c)) =  l(d)$ if
  $\major(a,b,c)  =  d$ for  the  tuples  $a,b,c,d  \in F^H_r$.   This
  extension remains a majority operation.

  The  relation~$\graph F$  is  obviously closed  under this  extended
  majority operation.  Jeavons  proved in~\cite{JeavonsCG-97} that the
  closure  of a  constraint  language~$S$ under  a majority  operation
  implies the  membership of  $\csp(S)$ in~$\p$, therefore  our result
  follows.
  %
\end{proof}

The       second        case       differs       from       Schaefer's
characterization~\cite{Schaefer-78}.   Before  introducing this  case,
let us remind  that a semi-lattice operation is  a binary associative,
commutative and idempotent operation.

\begin{proposition}\label{prop:order}
  Let~$F$ be a set of co-Boolean functions. If~$F^H_r$ is closed under
  conjunction  or disjunction, and  the first  two tuples  $(0,a)$ and
  $(1,b)$ of  the relation~$F^H$ satisfy the condition  $a \preceq b$,
  where~$\preceq$  is  the coordinate-wise  partial  order on  Boolean
  tuples  induced by  the  order $0  <  1$, then  $\csp(\graph F)$  is
  in~$\p$.
\end{proposition}
\begin{proof}
  The proof  is similar to  that of Proposition~\ref{prop:majorminor}.
  We    extend    the    operations    of    conjunction~$\land$    or
  disjunction~$\lor$   on  $\{0,1\}$   from~$F^H_r$   to  semi-lattice
  operations of~$F^H$  by using the  induced numbering of  tuples from
  the  first coordinate  of~$F^H$.  Jeavons~\cite{JeavonsCG-97} proved
  that the  closure of a constraint language~$S$  under a semi-lattice
  operation  implies   the  membership  of   $\csp(S)$  in~$\p$.  Then
  $\csp(F^H)$  is  in $\p$  and  therefore  also  $\csp(\graph F)$  by
  Proposition~\ref{prop:equiv}.
  %
\end{proof}

Finally we  show that the remaining  cases, when the  two first tuples
in~$F^H$  are   not  ordered  compatibly  with   their  position,  are
$\np$-complete.

\begin{proposition}\label{prop:nonorder}
  Let~$F$ be a set of co-Boolean functions. If~$F^H_r$ is closed under
  conjunction or disjunction, but neither under majority nor minority,
  and  the first  two tuples  $(0,a)$  and $(1,b)$  from~$F^H$ do  not
  satisfy  the  condition  $a  \preceq b$,  with~$\preceq$  being  the
  coordinate-wise partial order on Boolean tuples induced by the order
  $0 < 1$, then $\csp(\graph F)$ is $\np$-complete.
\end{proposition}
\begin{proof}
  Note  that~$F^H_r$  cannot  be  closed under  both  conjunction  and
  disjunction since this would imply that it is closed under majority.
  It follows  from the identity $\major(x,y,z)  = (x \lor  y) \land (y
  \lor z) \land (z \lor x)$.

  Consider the first two  tuples in~$F^H$, namely $(0,a)$ and $(1,b)$.
  The condition $a \preceq b$ is falsified, therefore there must exist
  a coordinate~$i$  with $a[i]=1$ and  $b[i]=0$. Hence there  exists a
  function $f \in F$ with  $f(0)=1$ and $f(1)=0$. Hence we have $\{01,
  10\}  \subseteq \graph  f$.   The relation  $\{01,  10\}$ is  closed
  neither  under conjunction, nor  under disjunction.   Therefore also
  the relation~$\graph  f$ cannot be closed under  minimum or maximum,
  hence  also the  relation~$F^H$  cannot be  closed  under these  two
  operations  either.  The Boolean  relation~$F^H_r$ is  by assumption
  closed neither under majority  nor under minority. Hence $\csp(F^H_r
  \cup     \graph    f)$     is     $\np$-complete    according     to
  Schaefer~\cite{Schaefer-78}.  Since the inclusion $F^H_r \cup \graph
  f  \subseteq  \ccln{F^H}$  holds,  by  Theorem~\ref{th:jeavons}  and
  Proposition~\ref{prop:equiv}     we     have    $\csp(\graph     F)$
  $\np$-complete.
  %
\end{proof}

By regrouping the proposition of this section, we derive the following
complete  classification  of  complexity  for  homogeneous  co-Boolean
$\csp$s.

\begin{theorem}[Dichotomy Theorem]\label{th:dicho}
  Let~$F$ be a set of  homogeneous co-Boolean functions and $\graph G$
  be a  set a relations such that  $\graph G = \core  \graph F$ holds.
  If  the  Boolean  relation~$G^H_r$   is  closed  under  majority  or
  minority, or  if it is  closed under conjunction or  disjunction and
  the two first tuples  $(0,a)$, $(1,b)$ from the matrix~$G^H$ satisfy
  the  condition $a  \preceq  b$, then  $\csp(\graph  F)$ is  in~$\p$.
  Otherwise $\csp(\graph F)$ is $\np$-complete.
\end{theorem}
\begin{proof}
  This     follows      from     Lemmata~\ref{lemma:coregraph}     and
  \ref{lem:corecst},       and       Propositions~\ref{prop:nonclosed}
  to~\ref{prop:nonorder}.
  %
\end{proof}

\section{Concluding Remarks}
\label{sec:conclusion}

The result  presented in this  short note is  a first step  toward the
study  of the  complexity of  $\csp$  over unary  functions.  We  have
proved  a   Dichotomy  Theorem  for  the   complexity  of  homogeneous
co-Boolean  constraint satisfaction problems  for every  finite domain
$D$.  Even  if this  Dichotomy Theorem is  mainly based on  the famous
Schaefer's  theorem, it presents  a first  study of  the new  angle of
attack  proposed by  Feder \textit{et  al.},  and allow  us to  easily
determinate the complexity of  homogeneous co-Boolean $\csp$ thanks to
the $H$-normal form, which was not trivial so far.

A natural extension of this work  would be the study of the complexity
of  $\csp(\graph  F)$  where~$F$  is  a set  of  unary  functions  not
necessary sharing anymore the same co-domain $\{0, 1\}$, \textit{i.e.}
non-homogeneous co-Boolean functions  where co-domains of functions $f
\in F$  are independent.   It would be  also interesting to  study the
complexity  of co-ternary  $\csp$,  where the  constraint language  is
built  upon  unary functions  sharing  the  same co-domain  $\{0,1,2\}
\subseteq D$.  Similarly to  our dichotomy theorem, which differs from
Schaefer's Dichotomy Theorem in~\cite{Schaefer-78}, we conjecture that
a dichotomy theorem for the latter problem will also be different from
Bulatov's Dichotomy Theorem in~\cite{Bulatov-06}.

\section{Acknowledgments}
The  author would  like to  thank Miki  Hermann and  Gustav  Nordh for
theirs helpful advices and ideas, allowing him to start this study.

\bibliographystyle{elsarticle-num}

\end{document}